\documentclass[times,10pt,twocolumn]{article}
\usepackage{times}
\usepackage{sty/latex8}
\usepackage[top=0.7in, bottom=0.7in, left=0.7in, right=0.7in]{geometry}
\usepackage{url}
\usepackage{xspace}
\usepackage[compact]{titlesec}
\usepackage{setspace}

\titleformat{\section}[hang]{\normalfont\large\bfseries}{\thesection. }{0pt}{}{}
\titleformat{\subsection}[hang]{\normalfont\bfseries}{\thesubsection. }{0pt}{}{}
\titleformat{\subsubsection}[runin]{\normalfont\bfseries}{\thesubsubsection. }{0pt}{}{}

\newlength{\sectionspace}
\newlength{\subsectionspace}
\titlespacing*{\section}{0pt}{\sectionspace}{\sectionspace}
\titlespacing*{\subsection}{0pt}{\subsectionspace}{\subsectionspace}
\titlespacing*{\subsubsection}{0pt}{\parskip}{0.2em plus 0.1em minus 0.05em}
\usepackage{footmisc}
\usepackage[normalem]{ulem}
\usepackage{float}
\usepackage{graphicx}
\usepackage{booktabs}
\usepackage{multirow}
\usepackage{subfig}
\usepackage{color}
\usepackage{xcolor}
\usepackage{pifont}
\usepackage[font={bf,footnotesize}]{caption}

\usepackage{gensymb}

\let\oldcelsius\celsius
\renewcommand{\celsius}{~\oldcelsius\xspace\xspace}
\newcommand{\ALD}[0]{AL-DRAM\xspace}

\newcommand{\squishlist}{
	\begin{list}{$\bullet$}
		{ \setlength{\itemsep}{0pt}      \setlength{\parsep}{3pt}
			\setlength{\topsep}{3pt}       \setlength{\partopsep}{0pt}
			\setlength{\leftmargin}{1.5em} \setlength{\labelwidth}{1em}
			\setlength{\labelsep}{0.5em} } }
\newcommand{\squishend}{
    \end{list}  }

\newcommand{\DIMMs}[0]{115\xspace}

\newcommand{\ReadHot}[0]{21.1}
\newcommand{\WriteHot}[0]{34.4}

\newcommand{\trcdHot}[0]{15.6}
\newcommand{\trasHot}[0]{20.4}
\newcommand{\twrHot}[0]{20.6}
\newcommand{\trpHot}[0]{28.5}

\newcommand{\ReadCold}[0]{32.7}
\newcommand{\WriteCold}[0]{55.1}

\newcommand{\trcdCold}[0]{17.3}
\newcommand{\trasCold}[0]{37.7}
\newcommand{\twrCold}[0]{54.8}
\newcommand{\trpCold}[0]{35.2}

\newcommand{\mycolor}[1]{\textcolor{black}{#1}}

\usepackage[numbers, sort]{natbib}
\usepackage{titling}

\usepackage{fancyhdr}
\fancypagestyle{plain}{
	\fancyhf{}
	\chead{SAFARI Technical Report No. 2016-003 (March 28, 2016)}
	\cfoot{\thepage}
	
}

\begin{document} 
\date{}
\pagestyle{plain}

\pretitle{\begin{center} \vskip -0.5in
This is a summary of the original paper, entitled ``Adaptive-Latency DRAM:
Optimizing DRAM Timing for the Common-Case'' which appears in HPCA
2015~\cite{lee-hpca2015}.\\ }

\title{\Large\bf
\vskip 0.15in
Adaptive-Latency DRAM (AL-DRAM)\\
\vskip -0.15in \vspace{0.0in}
}

\author{\vspace{-0.01in}
 	Donghyuk Lee \hspace{0.2in}
 	Yoongu Kim \hspace{0.2in}
 	Gennady Pekhimenko \\\vspace{0.05in}
 	Samira Khan \hspace{0.2in}
 	Vivek Seshadri \hspace{0.2in}
 	Kevin Chang \hspace{0.2in}
 	Onur Mutlu \\\vspace{0.03in}
 	Carnegie Mellon University\\ \vspace{-0.5in}
}

\maketitle

\setstretch{0.95}
\begin{abstract}

This paper summarizes the idea of Adaptive-Latency DRAM (AL-DRAM), which was
published in HPCA 2015~\cite{lee-hpca2015}. The key goal of AL-DRAM is to
exploit the extra margin that is built into the DRAM timing parameters to reduce
DRAM latency. The key observation is that the timing parameters are dictated by
the worst-case temperatures and worst-case DRAM cells, both of which lead to
small amount of charge storage and hence high access latency. One can therefore
reduce latency by \mycolor{adapting} the timing parameters to the current
operating temperature and the current DIMM that is being accessed. Using an
FPGA-based testing platform, \mycolor{our} work first characterizes the extra
margin for \DIMMs DRAM modules from three major manufacturers. The experimental
results demonstrate that it is possible to reduce four of the most critical
timing parameters by a minimum/maximum of \trcdCold\%/\twrCold\% at
55\celsius~\mycolor{while maintaining reliable operation.} AL-DRAM adaptively
selects between multiple different timing parameters for each DRAM module based
on its current operating condition. AL-DRAM does not require any changes to the
DRAM chip or its interface; it only requires multiple different timing
parameters to be specified and supported by the memory controller. Real system
evaluations show that AL-DRAM improves the performance of memory-intensive
workloads by an average of 14\% without introducing any
errors~\cite{lee-hpca2015}.

\end{abstract}

\section{Summary} \label{sec:summary}

\subsection{Problem: High DRAM Latency} \label{sec:problem}

A DRAM chip is made of capacitor-based cells that represent data in the form of
electrical charge. To store data in a cell, charge is injected, whereas to
retrieve data from a cell, charge is extracted. Such {\em movement of charge}
happens through a wire called {\em bitline}. Due to the large resistance and the
large capacitance of the bitline, it takes long time to access DRAM cells. To
guarantee correct operation, DRAM manufacturers impose a set of minimum latency
restrictions on DRAM accesses, called {\em timing parameters}~\cite{jedec-ddr3}.
Ideally, timing parameters should provide {\em just enough} time for a DRAM chip
to operate correctly. In practice, however, there is a very large margin in the
timing parameters to ensure correct operation under {\em worst-case} conditions
with respect to two aspects. First, due to {\em process variation}, some outlier
cells suffer from a larger RC-delay than other cells~\cite{lee-iedm1996,
kang-memforum2014}, and require more time to be accessed. Second, due to {\em
temperature dependence}, DRAM cells lose more charge at high
temperature~\cite{yaney-iedm1987}, and \mycolor{therefore} require more time to
be sensed and restored. Due to the worst-case \mycolor{provisioning} of timing
parameters, it takes longer time to access most of DRAM \mycolor{under most
operating conditions} than necessary for correct operation.

\subsection{Key Observations and Our Goal} \label{sec:observation}

{\bf Most DRAM chips do {\em not} contain the worst-case cell with the largest
latency.} Using an FPGA-based testing platform, we profile \DIMMs DRAM modules
and observe that the slowest cell, having the smallest amount of charge, for a
typical chip is still faster than that of the worst-case chip. We expose the
large margin built into {\em DRAM} timing parameters. In particular, we identify
four timing parameters that are the most critical during a DRAM access: {\tt
tRCD}, {\tt tRAS}, {\tt tWR}, and {\tt tRP}. At 55\celsius, we demonstrate that
the parameters can be reduced by an average of \trcdCold\%, \trasCold\%,
\twrCold\%, and \trpCold\% while still maintaining correctness.

\noindent{\bf Most DRAM chips are {\em not} exposed to the worst-case
temperature of 85\celsius.} We measure the DRAM ambient temperature in a server
cluster running a \mycolor{very} memory-intensive benchmark, and \mycolor{find}
that the temperature {\em never} exceeds 34\celsius\xspace\xspace --- as well as
never changing by more than 0.1\celsius\xspace per second. Other
works~\cite{elsayed-sigmetrics2012, elsayed-techreport2012, liu-hpca2011} also
observed that worst-case DRAM \mycolor{temperatures} are not common and servers
operate at much lower temperatures~\cite{elsayed-sigmetrics2012,
elsayed-techreport2012, liu-hpca2011}.

Based on these observations, we show {\em that} typical DRAM chips operating at
typical temperatures (e.g., 55\celsius) are capable of providing a much smaller
access latency, but are nevertheless forced to operate at the largest latency of
the worst-case due to the use of only a single set of timing parameters dictated
by the worst case.

{\bf Our goal} in our HPCA 2015 paper~\cite{lee-hpca2015} is to exploit the
extra margin that is built into the DRAM timing parameters to reduce DRAM
latency and thus improve performance. To this end, we first provide a detailed
analysis of why we can reduce DRAM timing parameters without sacrificing
reliability.

\subsection{Charge \& Latency Interdependence}

The operation of a DRAM cell is governed by two important parameters: {\em
i)}~the quantity of charge and {\em ii)}~the latency it takes to move charge.
These two parameters are closely related to each other. Based on SPICE
simulations with a detailed DRAM model, we identify the quantitative
relationship between charge and latency~\cite{lee-hpca2015}. While our HPCA 2015
paper provides the detailed analyses of this relationship, \mycolor{here} we
summarize the three key observations. First, having more charge in a DRAM cell
accelerates the sensing \mycolor{operation} in the cell, especially at the
beginning of sensing, enabling the opportunity to shorten the corresponding
timing parameters ({\tt tRCD} and {\tt tRAS}). Second, when restoring the charge
in a DRAM cell, a large amount of the time is spent on injecting the final small
amount of charge into the cell. If there is already enough charge in the cell
for the next access, the cell does not need to be fully restored. In this case,
it is possible to shorten the latter part of the restoration time, creating the
opportunity to shorten the corresponding timing parameters ({\tt tRAS} and {\tt
tWR}). Third, at the end of precharging, i.e., setting the bitline into the
initial voltage level (before accessing a cell) for the next access, a large
amount of the time is spent on precharging the final small amount of bitline
voltage difference from the initial level. When there is already enough charge
in the cell to overcome the voltage difference in the bitline, the bitline does
not need to be fully precharged. Thus, it is possible to shorten the final part
of the precharge time, creating the opportunity to shorten the corresponding
timing parameter ({\tt tRP}). Based on these three observations, we understand
that {\em timing parameters can be shortened if DRAM cells have enough charge}.

\subsection{Adaptive-Latency DRAM} \label{sec:observation}

As explained, the amount of charge in the cell right before an access to it
plays a critical role in whether the correct data is retrieved from the cell.
In Figure~\ref{fig:aldram}, we illustrate the impact of process variation using
two different cells: one is a {\em typical} cell (left column) and the other is
the worst-case cell which deviates the most from the typical (right column). The
worst-case cell contains less charge than the typical cell in its initial state.
This is because of two reasons. First, due to its {\em large resistance}, the
worst-case cell cannot allow charge to flow inside quickly. Second, due to its
{\em small capacitance}, the worst-case cell cannot store much charge even when
it is full. To accommodate such a worst-case cell, existing timing parameters
are set to a large value. 

In Figure~\ref{fig:aldram}, we also illustrate the impact of temperature
dependence using two cells at two different temperatures: {\em i)} a typical
temperature (55\celsius, bottom row), and {\em ii)} the worst-case temperature
(85\celsius, top row) supported by DRAM standards. Both typical and worst-case
cells leak charge at a faster rate at the worst-case temperature. Therefore, not
only does the worst-case cell have less charge to begin with, but it is left
with {\em even less} charge at the worst temperature because it leaks charge at
a faster rate (top-right in Figure~\ref{fig:aldram}). To accommodate the
combined effect of process variation {\em and} temperature dependence, existing
timing parameters are set to a very large value. That is why the worst-case
condition for correctness is specified by the top-right of
Figure~\ref{fig:aldram}, which shows the least amount of charge stored in {\em
the worst-case cell at the worst-case temperature} in its initial state. On top
of this, DRAM manufacturers \mycolor{still add an extra latency margin} even for
such worst-case conditions. In other words, the amount of charge at the
worst-case condition is still greater than what is required for correctness
under that condition.

\begin{figure}[h]
	\center
	\vspace{0.05in}
	\includegraphics[width=0.98\linewidth]{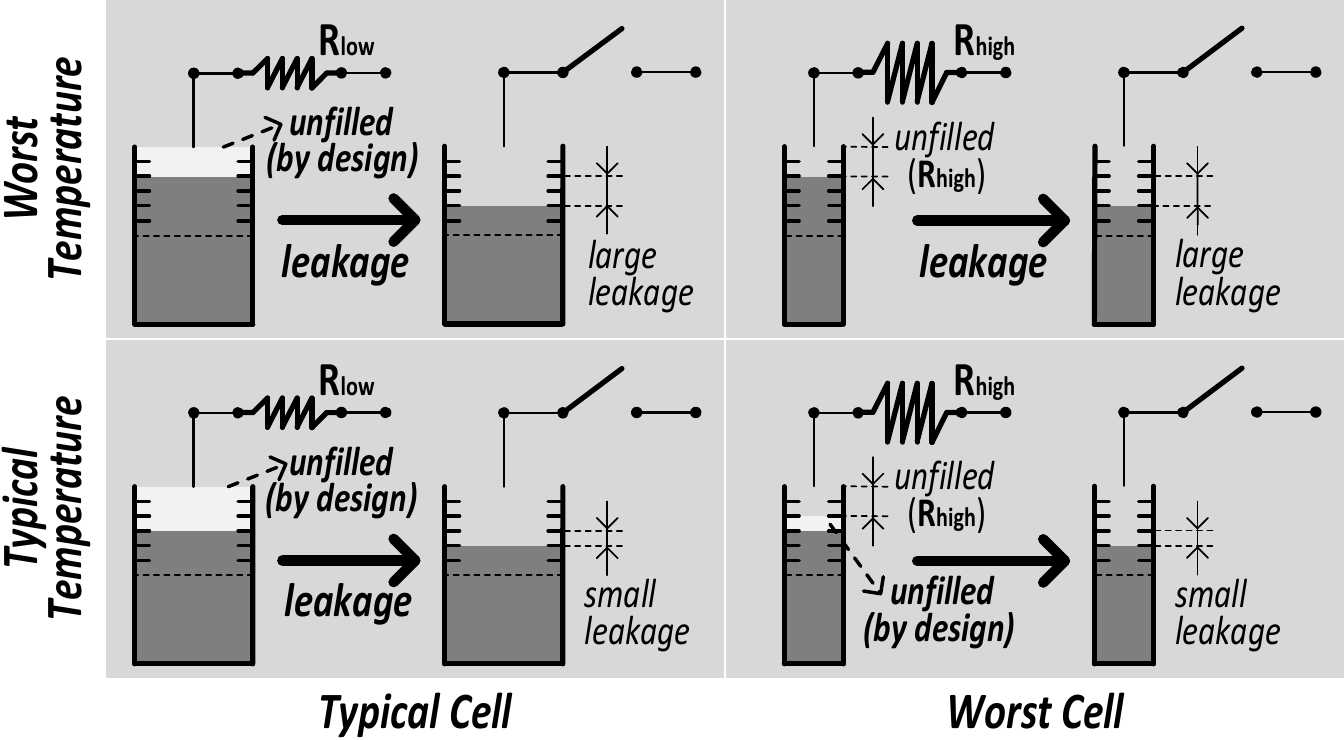}
	\caption{Effect of Reduced Latency: Typical~vs.~Worst}
	\label{fig:aldram}
	\vspace{0.10in}
\end{figure}

If we were to reduce the timing parameters, we would also be reducing the charge
stored in the cells. It is important to note, however, that we are proposing to
exploit {\em only} the {\em additional slack} (in terms of charge) compared to
the worst-case. This allows us to provide as strong of a reliability guarantee
as the worst-case. In Figure~\ref{fig:aldram}, we illustrate the impact of
reducing the timing parameters. The lightened portions inside the cells
represent the amount of charge that we are giving up by using the reduced timing
parameters. Note that we are not giving up any charge for the worst-case cell at
the worst-case temperature. Although the other three cells are not fully charged
in their initial state, they are left with a similar amount of charge as the
worst-case (top-right). This is because these cells are capable of either
holding more charge (typical cell, left column) or holding their charge longer
(typical temperature, bottom row). Therefore, optimizing the timing parameters
(based on the amount of existing charge slack) provides the opportunity to
reduce overall DRAM latency while still maintaining the reliability guarantees
provided by the DRAM manufacturers.

Based on these observations, we propose Adaptive-Latency DRAM (\ALD), a
mechanism that dynamically optimizes the timing parameters for different modules
at different temperatures. \ALD exploits the {\em additional charge slack}
present in the common-case compared to the worst-case, thereby preserving the
level of reliability (at least as high as the worst-case) provided by DRAM
manufacturers.

\subsection{DRAM Latency Profiling} \label{sec:profiling}

We present and analyze the results of our DRAM profiling experiments, performed
on our FPGA-based DRAM testing infrastructure~\cite{liu-isca2013,
khan-sigmetrics2014, kim-isca2014, qureshi-dsn2015, lee-hpca2015,
chang-sigmetrics2016, khan-dsn2016}. Figures~\ref{fig:read_latency}
and~\ref{fig:write_latency} show the results of this experiment for the read and
write latency tests. The y-axis plots the sum of the relevant timing parameters
({\tt tRCD}, {\tt tRAS}, and {\tt tRP} for the read latency test and {\tt tRCD},
{\tt tWR}, and {\tt tRP} for the write latency test). The solid black line shows
the latency sum of the standard timing parameters (DDR3 DRAM specification). The
dotted red line and the dotted blue line show the acceptable latency parameters
\mycolor{that do not cause any errors} for each DIMM at 85\celsius\xspace and
55\celsius, respectively. The solid red line and blue line show the average
acceptable latency across all DIMMs.

\begin{figure}[h]
	\vspace{-0.15in}
	\centering
	\subfloat [Read Latency] {
		\includegraphics[width=1.65in]{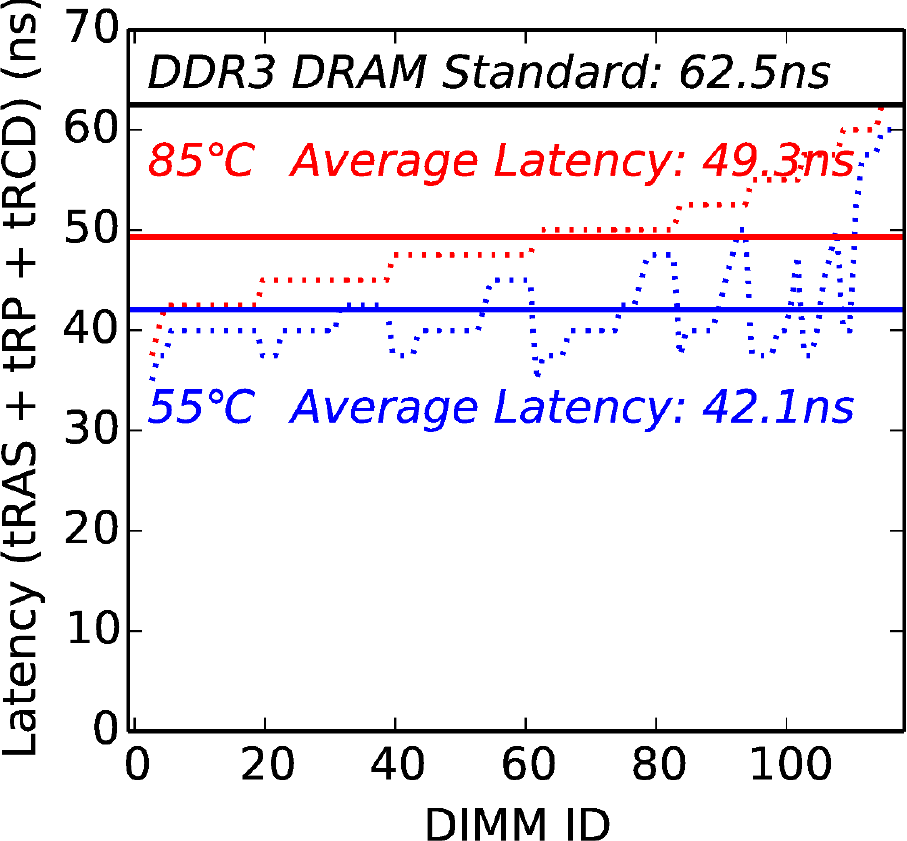}
		\label{fig:read_latency}
	}
	\subfloat [Write Latency] {
		\includegraphics[width=1.65in]{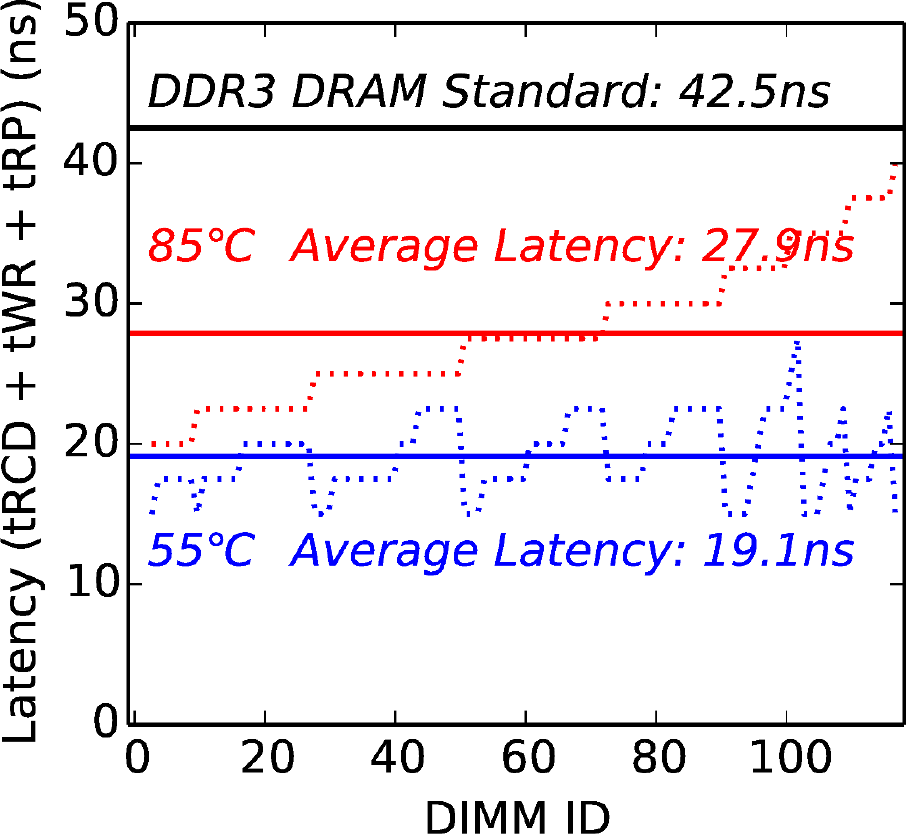}
		\label{fig:write_latency}
	}
	\caption{Access Latency Analysis of 115 DIMMs} \label{fig:multiple_dimms}
	\vspace{-0.05in}
\end{figure}

We make two observations. First, even at the highest temperature of 85\celsius,
DIMMs have a high potential of access with reduced latency: \ReadHot\% on
average for read, and \WriteHot\% on average for write operations. This is a
direct result of the possible reductions in timing parameters {\tt tRCD}/{\tt
tRAS}/{\tt tWR}/{\tt tRP} --- \trcdHot\%/\trasHot\%/\twrHot\%/\trpHot\% on
average across all the DIMMs. As a result, we conclude that process variation
and lower temperatures enable a significant potential to reduce DRAM access
latencies. Second, we observe that at lower temperatures (e.g., 55\celsius), the
potential for latency reduction is even greater (\ReadCold\% on average for
read, and \WriteCold\% on average for write operations), where the corresponding
reduction in timing parameters {\tt tRCD}/{\tt tRAS}/{\tt tWR}/{\tt tRP} are
\trcdCold\%/\trasCold\%/\twrCold\%/\trpCold\% on average across all the DIMMs.

\subsection{Real-System Evaluation} \label{sec:evaluation}

We evaluate \ALD on a real system that offers dynamic software-based control
over DRAM timing parameters at runtime~\cite{amd-4386, amd-bkdg}. We use the
minimum values of the timing parameters that do not introduce any errors at
55\celsius~for any module to determine the latency reduction at 55\celsius.
Thus, the latency is reduced by 27\%/32\%/33\%/18\% for {\tt tRCD}/{\tt
tRAS}/{\tt tWR}/{\tt tRP}, respectively. Our full methodology is described in
our HPCA 2015 paper~\cite{lee-hpca2015}.

Figure~\ref{fig:result_1r1c} shows the performance improvement of reducing the
timing parameters in the evaluated memory system with one rank and one memory
channel at 55\celsius~operating temperature. We run a variety of different
applications in two different configurations. The first one (single-core) runs
only one thread, and the second one (multi-core) runs multiple
applications/threads. We run each configuration 30 times (only SPEC
\mycolor{benchmarks are} executed 3 times due to \mycolor{their large execution
times}), and present the average performance improvement across all the runs and
their standard deviation as an error bar. Based on the last-level cache misses
per kilo instructions (MPKI), we categorize our applications into
memory-intensive or non-intensive groups, and report the geometric mean
performance improvement across all applications from each group. 

We draw three key conclusions from Figure~\ref{fig:result_1r1c}. First, \ALD
provides significant performance improvement over the baseline (as high as
20.5\% for the very memory-bandwidth-intensive STREAM
applications~\cite{moscibroda-usenix2007}). Second, when the memory system is
under higher pressure with multi-core/multi-threaded applications, we observe
significantly higher performance (than in the single-core case) across all
applications from our workload pool. Third, as expected, memory-intensive
applications benefit more in performance than non-memory-intensive workloads
(14.0\% vs.~2.9\% on average). We conclude that by reducing the DRAM timing
parameters using AL-DRAM, we can speed up a real system by 10.5\% (on average
across all 35 workloads on the multi-core/multi-thread configuration).

\begin{figure}[h]
	\centering
	\includegraphics[width=0.99\linewidth]{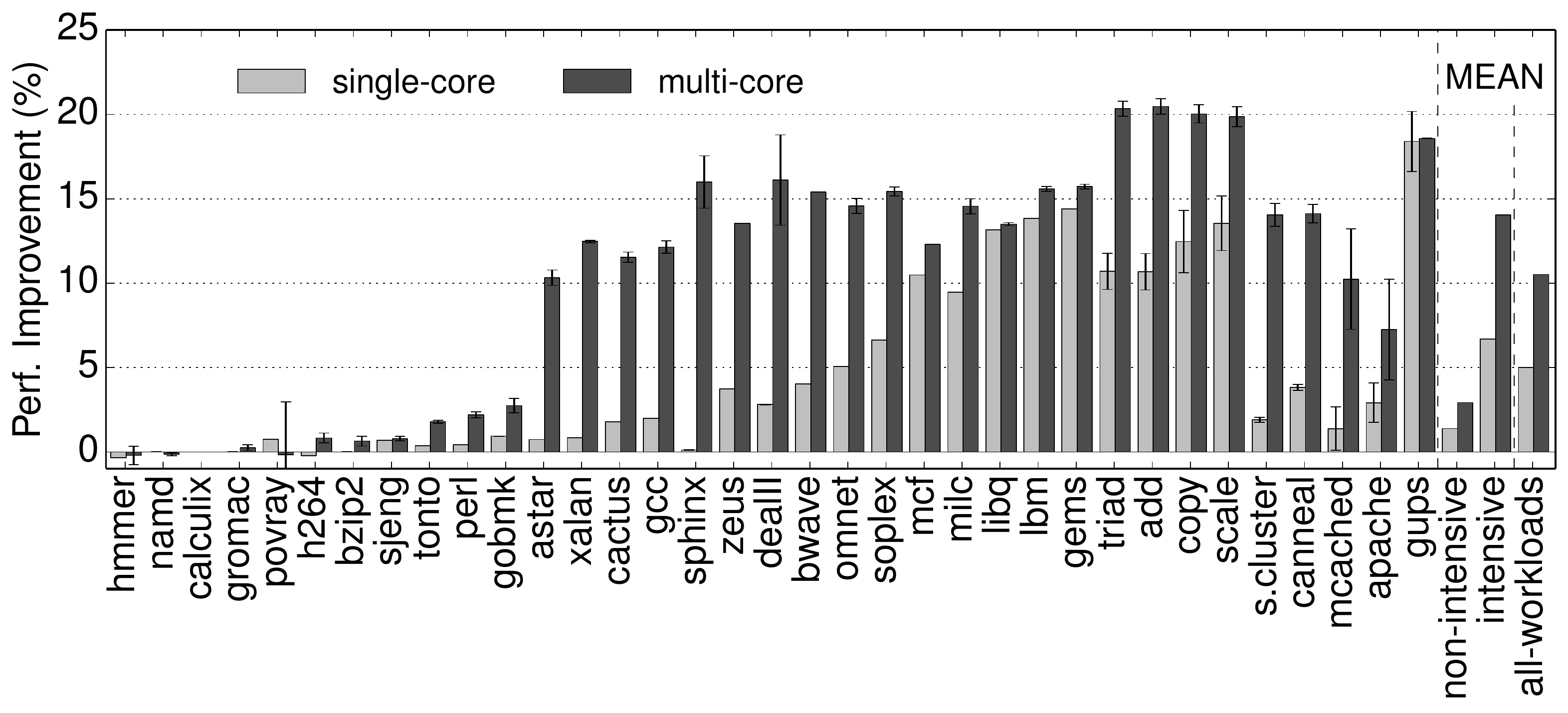}
	\captionof{figure}{Real System Performance Improvement with AL-DRAM}
	\label{fig:result_1r1c}
	\vspace{0.05in}
\end{figure}

\subsection{Other Results and Analyses in Our Paper} \label{sec:other_results}

Our HPCA paper includes more DRAM latency analyses and system performance
evaluations.

\squishlist

	\item {\bf Effect of Changing the Refresh Interval on DRAM Latency.} We
	evaluate DRAM latency at different refresh intervals. We observe that
	refreshing DRAM cells more frequently enables more DRAM latency reduction.

	\item {\bf Effect of Reducing Multiple Timing Parameters.} We study the
	potential for reducing multiple timing parameters simultaneously. Our key
	observation is that reducing one timing parameter leads to decreasing the
	opportunity to reduce another timing parameter simultaneously.

	\item {\bf Analysis of the Repeatability of Cell Failures.} We perform
	\mycolor{tests} for five different scenarios \mycolor{to determine that a cell
	failure due to reduced latency is repeatable:} same test, test with different
	data patterns, test with timing-parameter combinations, test with different
	temperatures, and read/write test. Most of these scenarios show that a very
	high fraction (more than 95\%) of the erroneous cells consistently experience
	an error over multiple iterations of the same test.

	\item {\bf Performance Sensitivity Analyses.} We analyze the impact of
	increasing the number of ranks and channels, executing heterogeneous
	workloads, using different \mycolor{row buffer policies}.

\squishend

\setstretch{0.963}
\section{Significance} \label{sec:significance}

\subsection{Novelty} \label{sec:novelty}

To our knowledge, our HPCA 2015 paper is the first work to {\em i)} provide a
detailed qualitative and empirical analysis of the relationship between {\em
process variation} and {\em temperature dependence} of modern DRAM devices on
the one side, and DRAM access latency on the other side (we directly attribute
the relationship between the two to {\em the amount of charge} in cells), {\em
ii)} experimentally characterize a large number of existing DIMMs to understand
the potential of reducing DRAM timing constraints, {\em iii)} provide a
practical mechanism that can take advantage of this potential, and {\em iv)}
evaluate the performance benefits of this mechanism by {\em dynamically
optimizing} DRAM timing parameters on a real system using a variety of real
workloads. We make the following major contributions.

{\bf Addressing a Critical Real Problem, High DRAM Latency, with Low Cost.} High
DRAM latency is a critical bottleneck for overall system performance in a
variety of modern computing systems~\cite{mutlu-superfri2015, mutlu-imw2013},
especially in real large-scale server systems~\cite{lo-isca2015}. Considering
the difficulties in DRAM scaling~\cite{mutlu-superfri2015, mutlu-imw2013,
kang-memforum2014}, the problem is getting worse in future systems due to
process variation. Our HPCA 2015 work leverages the heterogeneity created by
DRAM process variation \mycolor{across DRAM chips} and system operating
conditions to mitigate the DRAM latency problem. We propose a practical
mechanism, {\em Adaptive-Latency DRAM}, which mitigates DRAM latency with very
modest hardware cost, and with {\em no changes} to the DRAM chip itself.

{\bf Low Latency DRAM Architectures.} Previous works~\cite{mutlu-memcon2013,
sato-vlsi1998, rldram, lee-hpca2013, son-isca2013, kim-isca2012,
seshadri-micro2013, hidaka-ieeemicro1990, zhang-ieeemicro2001, chang-hpca2014,
chang-hpca2016} propose new DRAM architectures that provide lower latency.
These works improve DRAM latency at the cost of either significant additional
DRAM chip area (i.e., extra sense amplifiers~\cite{sato-vlsi1998, rldram,
son-isca2013}, an additional SRAM cache~\cite{hidaka-ieeemicro1990,
zhang-ieeemicro2001}), specialized protocols~\cite{seshadri-micro2013,
kim-isca2012, lee-hpca2013, chang-hpca2014} or \mycolor{a combination} of these.
Our proposed mechanism requires {\em no changes} to the DRAM chip and the DRAM
interface, and hence has almost negligible overhead. Furthermore, \ALD is
largely orthogonal to these proposed designs, and can be applied in conjunction
with them, providing greater cumulative reduction in latency.

{\bf Large-Scale Latency Profiling of Modern DRAM Chips.} Using our FPGA-based
DRAM testing infrastructure~\cite{liu-isca2013, khan-sigmetrics2014,
kim-isca2014, qureshi-dsn2015, lee-hpca2015, khan-dsn2016,
chang-sigmetrics2016}, we profile 115 DRAM modules (862 DRAM chips in total) and
show that there is significant timing variation between different DIMMs at
different temperatures. We believe that our results are statistically
significant to validate our hypothesis that the DRAM timing parameters strongly
depend on the amount of cell charge. We provide detailed characterization of
each DIMM online at the SAFARI Research Group website~\cite{safari}.
Furthermore, we introduce our FPGA-based DRAM infrastructure and experimental
\mycolor{methodology} for DRAM profiling, which are carefully constructed to
represent the worst-case conditions in power noise, bitline/wordline coupling,
data patterns, and access patterns. Such information will hopefully be useful
for future DRAM research.

{\bf Extensive {\em Real} System Evaluation of DRAM Latency.} We evaluate our
mechanism on a real system and show that our mechanism provides significant
performance improvement. Reducing the timing parameters strips the excessive
margin in DRAM's electrical charge. We show that the remaining margin is {\em
enough} for DRAM to operate correctly. To verify the correctness of our
experiments, we ran our workloads for 33 days non-stop, and examined their and
the system's correctness with reduced timing parameters. Using the reduced
timing parameters over the course of 33 days, our real system was able to
execute 35 different workloads in both single-core and multi-core configurations
while preserving correctness and being {\em error-free}. Note that these results
do {\em not} absolutely guarantee that no errors can be introduced by reducing
the timing parameters. However, we believe that we have demonstrated a
proof-of-concept which shows that DRAM latency can be reduced at \mycolor{no}
impact on DRAM reliability. Ultimately, the DRAM manufacturers can provide the
reliable timing parameters for different operating conditions and modules.

{\bf Other Methods for Lowering Memory Latency.} There are many works that
reduce {\em overall memory access latency} by modifying DRAM, the
DRAM-controller interface, and DRAM controllers. These works enable more
parallelism and bandwidth~\cite{kim-isca2012, chang-hpca2014,
seshadri-micro2013, lee-taco2016, chang-hpca2016, zhang-isca2014, ahn-taco2012,
ahn-cal2009, ware-iccd2006, zheng-micro2008}, reduce refresh
counts~\cite{liu-isca2012, liu-isca2013, khan-sigmetrics2014,
venkatesan-hpca2006, qureshi-dsn2015}, accelerate bulk
operations~\cite{seshadri-micro2013, seshadri-cal2015, seshadri-micro2015,
chang-hpca2016}, accelerate computation in the logic layer of 3D-stacked
DRAM~\cite{ahn-isca2015a, ahn-isca2015b, zhang-hpca2014, guo-wondp2014}, enable
better communication between CPU and other devices through
DRAM~\cite{lee-pact2015}, leverage DRAM access patterns~\cite{hassan-hpca2016},
reduce write-related latencies by better designing DRAM and DRAM control
policies~\cite{chatterjee-hpca2012, lee-techreport2010, seshadri-isca2014},
reduce overall queuing latencies in DRAM by better scheduling memory
requests~\cite{rixner-isca2000, moscibroda-podc2008, lee-micro2009,
nesbit-micro2006, moscibroda-usenix2007, ausavarungnirun-isca2012,
ausavarungnirun-pact2015, zhao-micro2014, mutlu-micro2007, mutlu-isca2008,
ebrahimi-micro2011, kim-hpca2010, kim-micro2010, das-hpca2013,
muralihara-micro2011, jog-sigmetrics2016, subramanian-tpds2016,
subramanian-iccd2014, ipek-isca2008, usui-taco2016, subramanian-micro2015,
subramanian-hpca2013, lee-micro2008, lee-tc2011}, employing
prefetching~\cite{srinath-hpca2007, patterson-sosp1995, nesbit-pact2004,
ebrahimi-hpca2009, ebrahimi-micro2009, ebrahimi-isca2011, dahlgren-tpds1995,
alameldeen-hpca2007, cao-sigmetrics1995, lee-micro2008, mutlu-hpca2003,
mutlu-ieeemicro2003, mutlu-isca2005, mutlu-micro2005, dundas-ics1997,
cooksey-asplos2002, mutlu-ieeemicro2006}, memory/cache
compression~\cite{pekhimenko-micro2013, pekhimenko-pact2012, shafiee-hpca2014,
zhang-asplos2000, wilson-atec1999, dusser-ics2009, douglis-usenix1993,
decastro-sbacpad2003, alameldeen-tech2004, alameldeen-isca2004, abali-ibm2001,
pekhimenko-hpca2015, pekhimenko-hpca2016}, or better
caching~\cite{seshadri-pact2012, khan-hpca2014, qureshi-isca2007,
qureshi-isca2006}. Our proposal is orthogonal to all of these approaches and can
be applied in conjunction with them to achieve even higher latency reductions.

\subsection{Potential Long-Term Impact} \label{sec:longterm}

{\bf Tolerating High DRAM Latency by Exploiting DRAM Intrinsic Characteristics.}
Today, there is a large latency cliff between the on-chip last level cache and
off-chip DRAM, leading to a large performance fall-off when applications start
missing in the last level cache. By enabling lower DRAM latency, our mechanism,
Adaptive-Latency DRAM, smoothens this latency cliff \mycolor{without} adding
another layer into the memory hierarchy.

{\bf Applicability to Future Memory Devices.} We show the benefits of the
common-case timing optimization in modern DRAM devices by taking advantage of
intrinsic characteristics of DRAM. Considering that most memory devices adopt a
unified specification that is dictated by the worst-case operating condition,
our approach that optimizes device latency for the common case can be applicable
to other memory devices by leveraging the intrinsic characteristics of the
technology they are built with. We believe there is significant potential for
approaches that could reduce the latency of Phase Change Memory
(PCM)~\cite{raoux-ibm2008, lee-isca2009, qureshi-isca2009, qureshi-micro2009,
dhiman-dac2009, lee-ieeemicro2010, lee-cacm2010, meza-iccd2012, yoon-taco2014,
wong-ieee2010}, STT-MRAM~\cite{kultursay-ispass2013, meza-iccd2012,
li-arxiv2015}, RRAM~\cite{wong-ieee2012}, and Flash memory~\cite{luo-msst2015,
cai-date2013, cai-itj2013, cai-iccd2012, cai-sigmetrics2014, cai-iccd2013,
cai-dsn2015, cai-hpca2015, meza-sigmetrics2015, lu-tc2015, cai-date2012}.

{\bf New Research Opportunities.} Adaptive-Latency DRAM creates new
opportunities by enabling mechanisms that can leverage the heterogeneous latency
offered by our mechanism. We describe a few of these briefly.

{\em Optimizing the operating conditions for faster DRAM access.}
Adaptive-Latency DRAM provides different access latency at different operating
conditions. Thus, optimizing the DRAM operating conditions enables faster DRAM
access with Adaptive-Latency DRAM. For instance, balancing DRAM accesses over
different DRAM channels and ranks leads to reducing the DRAM operating
temperature, maximizing the benefits from Adaptive-Latency DRAM. At the system
level, operating the system at a constant low temperature can enable the use of
AL-DRAM's lower latency more frequently.

{\em Optimizing data placement for reducing overall DRAM access latency.} We
characterize the latency variation in different DIMMs due to process variation.
Placing data based on this information and the latency criticality of data
maximizes the benefits of lowering DRAM latency.

{\em Error-correction mechanisms to further reduce DRAM latency.}
Error-correction mechanisms can fix the errors from lowering DRAM latency even
further, leading to further reduction in DRAM latency without errors. Future
research that uses error correction to enable \mycolor{even} lower latency DRAM
is therefore promising as it opens a new set of trade-offs.

\bibliographystyle{sty/abbrv_etal}

\setstretch{0.955}
\footnotesize{
	\bibliography{paper}
}

\end{document}